\documentclass[twoside,twocolumn,english,superscriptaddress,nofootinbib,preprintnumbers, aps]{revtex4}
\usepackage{amsmath,amsthm,amssymb,amsfonts}
\usepackage[unicode=true, pdfusetitle, bookmarks=true,bookmarksnumbered=false,bookmarksopen=false, breaklinks=false,pdfborder={0 0 1},backref=false,colorlinks=false]{hyperref}
\usepackage{MnSymbol}
\usepackage{colordvi}
\usepackage{color} 
\usepackage{graphicx}
\usepackage[lofdepth,lotdepth]{subfig}
\usepackage[usenames,dvipsnames]{xcolor}
\usepackage[T1]{fontenc}
\usepackage[latin9]{inputenc}

\numberwithin{equation}{section}
\numberwithin{thr}{section}
\numberwithin{chr}{section}
\numberwithin{df}{section}

\begin{document}

\title{Relational evolution of observables for Hamiltonian-constrained systems}

\author{Andrea Dapor}
\email{adapor@fuw.edu.pl} \affiliation{Faculty of Physics, University of Warsaw, Ho\.{z}a 69, 00-681 Warszawa, Poland}

\author{Wojciech Kami\'nski}
\email{wojciech.kaminski@fuw.edu.pl} \affiliation{Faculty of Physics, University of Warsaw, Ho\.{z}a 69, 00-681 Warszawa, Poland}

\author{Jerzy Lewandowski}
\email{jerzy.lewandowski@fuw.edu.pl} \affiliation{Faculty of Physics, University of Warsaw, Ho\.{z}a 69, 00-681 Warszawa, Poland}
\affiliation{Institute for Quantum Gravity (IQG), FAU Erlangen -- Nurnberg, Staudtstr. 7, 91058 Erlangen, Germany}

\author{J\k{e}drzej \'Swie\.zewski}
\email{swiezew@fuw.edu.pl} \affiliation{Faculty of Physics, University of Warsaw, Ho\.{z}a 69, 00-681 Warszawa, Poland}


\begin{abstract}
Evolution of systems in which Hamiltonians are generators of gauge 
transformations is a notion that requires more structure than the 
canonical theory provides. We identify and study this additional 
structure in the framework of relational observables
(``partial  observables'').  We formulate necessary and sufficient conditions 
for the resulting evolution in the physical phase space to be a 
symplectomorphism. We give examples which satisfy those conditions and 
examples which do not. We point out that several classic positions in the 
literature on relational observables contain an incomplete approach to the issue
of evolution and false statements. Our work provides useful clarification and opens
the door to studying correctly formulated definitions.     

\end{abstract}
      
\date{\today}

\maketitle

\section{Introduction: relational Dirac observables}

This paper is about physical systems in which the Hamiltonian is a 
generator of a gauge transformation. The most important example is
general relativity. The physical evolution of those systems is a notion that requires
more care than in a regular canonical theory and an additional structure. On one hand, theoretical physicists dealing with such systems usually  know some methods which work in the examples they are interested in. On the other hand, there is a diversity of formulations of frameworks for the Hamiltonian-constrained systems \cite{cham1,cham2,cham3,cham4,cham5}, each one 
aiming to capture the peculiarity of this gauge-time evolution in a general way. The formulation we consider in this 
work originates from Rovelli's idea of ``partial observables'' \cite{rove}. We 
prefer to call it ``relational observables.'' A systematic approach to the relational observables framework was developed in a series of papers and  books \cite{Bianca1,Bianca2,tom,book}.  Unfortunately, we found an error in those attempts  \cite{Bianca1}. The error has passed to the literature and still requires a correction.  We would like to emphasise that, in specific examples, the idea of the relational observables is usually applied in a correct way
\cite{Bianca3}.  It is the general theory that needs our correction.
We present it in our current paper.

\subsubsection{Kinematical phase space}

In the (classical) canonical framework, the states form a phase space $\Gamma$. The phase space is a manifold endowed with a differential 2-form $\Omega$ --  a symplectic form -- which by definition is closed,
\begin{equation} d\Omega\  =\  0,  \end{equation}
and not degenerate,  
\begin{equation} X \righthalfcup\Omega\ =\ 0\ \Rightarrow\ X\ =\ 0,   \end{equation}
at every point of $\Gamma$.

\subsubsection{Constrained systems and gauge transformations}

In the case of a constrained system, 
the physical states are subject to constraints, and they form a constraint surface $\Gamma_C$ contained in the phase space 
\begin{equation} \Gamma_C\subset \Gamma. \end{equation}
On the constraint surface, there is a naturally induced 2-form $\Omega_C$, the pullback of $\Omega$. Provided $\Omega_C$ is nondegenerate, the pair $\Gamma_C$ and $\Omega_C$ becomes the physical phase space.  Often, however,     
the 2-form $\Omega_C$ has null directions, i.e., there is a nonvanishing vector  $\ell$ tangent to $\Gamma_C$, such that
\begin{equation} \ell\righthalfcup\Omega_{C}\ =\ 0.  \end{equation}
In this case $\Omega_C$ is degenerate; it cannot be a symplectic form, and 
$\Gamma_C$ cannot play the role of a physical phase
space. The maps $\Gamma_C\rightarrow\Gamma_C$ which define the flow of a null vector 
field $\ell$ are considered gauge transformations. The flow of each null vector field Lie-drags the 2-form $\Omega_C$, 
\begin{equation} {\cal L}_\ell\Omega_C \ =\ 0,\end{equation}
so $\Omega_C$ is gauge invariant.

\subsubsection{Physical phase space}

If $\ell_1$ and $\ell_2$ are two null vector fields, then so is their commutator,
\begin{equation}
[\ell_1,\ell_2]\righthalfcup\Omega_{C}\ =\ {\cal L}_{\ell_1}(\ell_2\righthalfcup\Omega_C)-\ell_2\righthalfcup{\cal L}_{\ell_1}\Omega_C\ =\ 0. 
\end{equation}
Therefore, the null directions define a foliation of $\Gamma_C$. The set of the leaves of that foliation is the physical phase space $\bar{\Gamma}$. We have the natural projection
\begin{equation}
\Pi:\Gamma_C\rightarrow \bar{\Gamma}.
\end{equation}             
We will be assuming that $\bar{\Gamma}$ is a manifold and that the natural 
projection is smooth. Then, there is a 2-form $\bar{\Omega}$ on $\bar{\Gamma}$
such that 
\begin{equation}
\Pi^*\bar{\Omega}\ =\ \Omega_C .
\end{equation}  
We will be also assuming that there is a global section of the projection $\Pi$, that is an embedding
\begin{equation}
\sigma:\bar{\Gamma}\rightarrow \Gamma_C,
\end{equation} 
such that combined with the natural projection $\Pi$, it is the identity
\begin{equation} \Pi\circ\sigma\ =\ {\rm id}. \end{equation}
In other words, our considerations will be local and will concern generic points. 
We will not address the issues of possible singular, or even non-Hausdorf points of $\Bar{\Gamma}$, or possible topological nontriviality of the projection $\Pi$.

The pair $(\bar{\Gamma},\bar{\Omega})$ is the physical phase space. Its points correspond to the physical states, and $\bar{\Omega}^{-1}$ defines the physical 
Poisson bracket: 
\begin{equation}\label{physPoiss} \{\bar{f},\bar{g}\}_{\rm phys}\ =\ (\bar{\Omega}^{-1}){}^{IJ}\partial_I\bar{f}\partial_J\bar{g}. \end{equation}

\begin{figure}[h]
\includegraphics[width=0.23\textwidth]{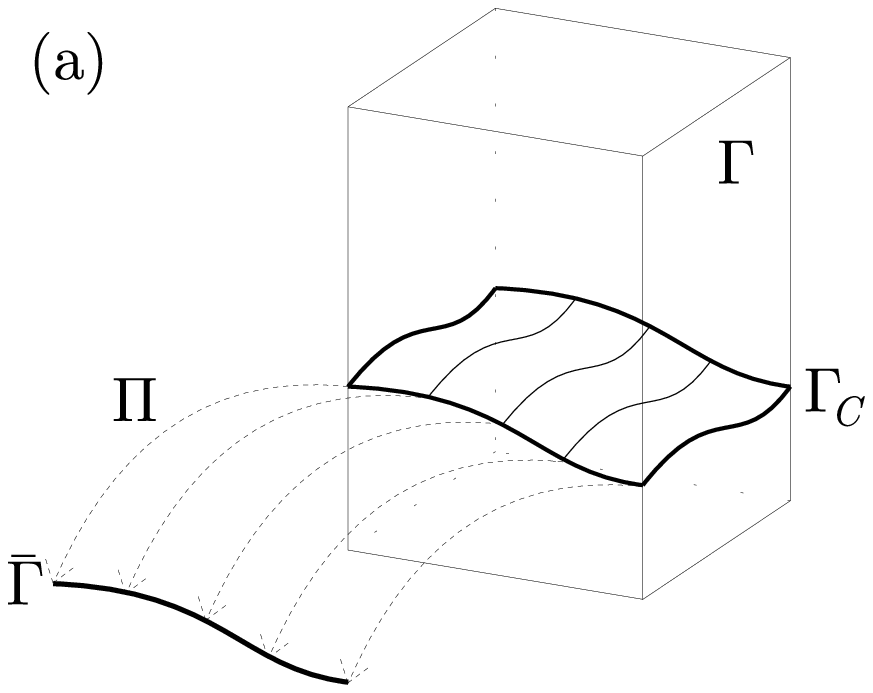}
\includegraphics[width=0.23\textwidth]{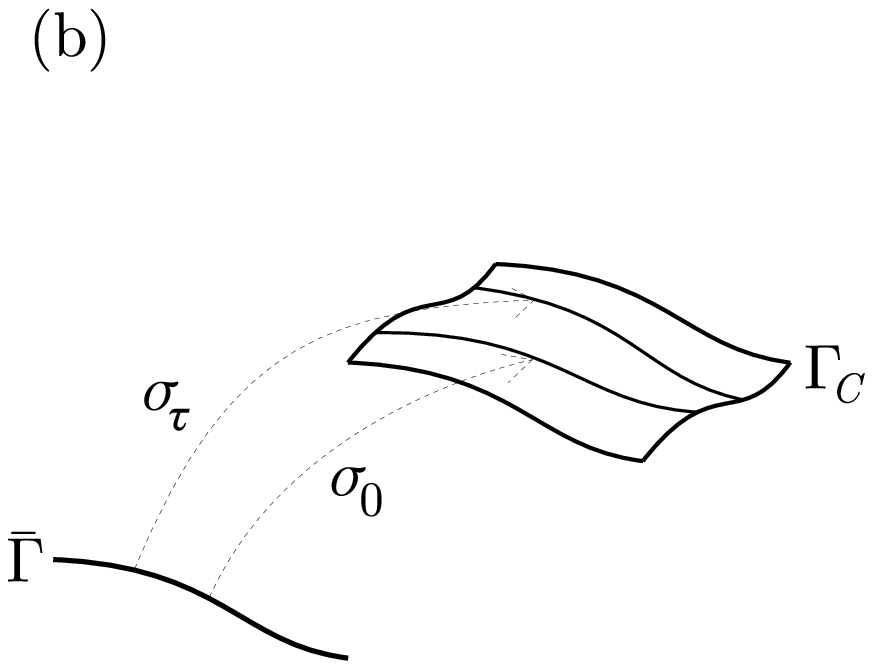}
\caption{The pictures illustrate the construction presented in the introduction. The picture (a) shows the phase space, the constraint surface, orbits of gauge transformations, the physical phase space, and the projection onto it.  The picture (b) shows the physical phase space, the constraint surface, and the ($\tau$-dependent) sections which embed one into the other. \label{pictures}}
\end{figure}

\subsubsection{Hamiltonians}

The dynamics of a canonical theory is defined by a Hamiltonian, a parameter $t$-dependent function $H(t)$ defined on the phase space $\Gamma$.  
At every value of $t$, the Hamiltonian $H(t)$ defines a vector field $X_{H(t)}$ tangent to $\Gamma$, such that 
\begin{equation}  X_{H(t)}\righthalfcup\Omega\ =\ dH(t).   \end{equation}
For every function $F$ defined on $\Gamma$, the Hamiltonian defines its evolution
$t\mapsto F(t)$ via
\begin{equation}\frac{d}{dt}F(t)\ =\ X_{H(t)}(F(t)),\ \ \ \ F(t_0)\ =\ F.\end{equation}
In the case of  constraints, the vector field $X_{H(t)}$ is tangent to $\Gamma_C$ at every instant of $t$, and its flow preserves the $\Omega_C$ including its null directions. The Hamiltonian $H(t)$ is constant along the leaves 
of the null directions,
\begin{equation} \ell(H(t))\ =\ \ell\righthalfcup dH(t)\ =\ \ell\righthalfcup(X_{H(t)}\righthalfcup \Omega)\ =\ 0. \end{equation}
Therefore,  the projection $\Pi_*X_{H(t)}$ defines a unique vector field   $X_{\bar{H}(t)}$ on $\bar{\Gamma}$, $H(t)$ defines a unique function $\bar{H}(t)$
on $\bar{\Gamma}$, and 
\begin{equation} X_{\bar{H}(t)}\righthalfcup\bar{\Omega}\ =\ d\bar{H}(t). \end{equation}
In fact, in a constrained system, the Hamiltonian vector field $X_{H(t)}$
is defined on $\Gamma_C$ modulo a null vector field by a nonunique Hamiltonian (that explains the plural ``Hamiltonians''), but its projection onto 
the physical phase space $\bar{\Gamma}$ is unique.
As a consequence, upon those smoothness assumptions, we end up with the physical phase space, symplectic form, and, respectively, Hamiltonian: $\bar{\Gamma},\  \bar{\Omega},\  \bar{H}(t)$.

\subsubsection{Hamiltonian-constrained systems}

There is a catch, however. In some constrained systems, the Hamiltonian vector field  $X_{H(t)}$ satisfies at every point of $\Gamma_C$,
\begin{equation} X_{H(t)} \righthalfcup \Omega\ =\ 0. \end{equation}
Then, the projected Hamiltonian vector field  is identically zero: 
\begin{equation}
X_{\bar{H}(t)}\ =\ 0.
\end{equation}
A system with this property is called Hamiltonian-constrained. An example is canonical general relativity, in which Hamiltonians (labelled by free functions -- lapse and shift -- representing the choice of space-time coordinates) are projected to $0$ in ${\bar{\Gamma}}$.

\subsubsection{Relational Dirac observables}

In order to introduce dynamics in the case of a Hamiltonian-constrained system, one needs some extra structure. Sometimes this extra structure is incorrectly characterized.We explain now this misunderstanding and identify the correct structure. We start by introducing the original construction
used in the literature. Consider a global section of the natural projection $\Pi$,
\begin{equation} \sigma_0 : \bar{\Gamma} \rightarrow \Gamma_0 \subset \Gamma_C \end{equation}
where we denoted $\Gamma_0 := \sigma_0(\bar{\Gamma})$. Every function $f$ defined
on $\Gamma_C$ determines a function $\bar{f}$ defined on $\bar{\Gamma}$
by the restriction of $f$ to the slice $\Gamma_0$,
\begin{equation} \label{sigmaf}\bar{f}\ :=\ \sigma_0^*f. \end{equation}
Suppose $T$ is a family of functions $T^m, m\in {\cal M}$, on $\Gamma_C$ such that the slice $\Gamma_0$ is defined as their common zero set
\begin{equation} \Gamma_0\ =\ \{\gamma\in \Gamma_C\ :\ T^m(\gamma)=0,\ m\in{\cal M}\}. \end{equation}
We can then consider any other slice defined by
\begin{equation}\label{sigmatau} T^m\ =\ \tau^m\in\mathbb{R}; \end{equation}
that is,
\begin{equation}\label{generalslice} \Gamma_\tau\ :=\ \{\gamma \in \Gamma_C : T^m(\gamma) = \tau^m, m \in {\cal M}\}. \end{equation}
In this way, for every family $\tau$ of numbers $\tau^m \in {\cal T}^m := (\tau_0^m, \tau_1^m) \subset \mathbb{R}$, the clock system $T$ defines a global section $\sigma_\tau$. Functions $T^m$ are called in the literature clock functions. Then, for every $\tau$ we obtain from a function $f$ a function on $\bar{\Gamma}$; generalizing Eq. (\ref{sigmaf}), it is
\begin{equation} \bar{f}_\tau\ :=\ \sigma_\tau^*f.\label{defsigmatau} \end{equation}
On the other hand, every function $\bar{f}$ on $\bar{\Gamma}$ can be equivalently expressed by
the function  $\Pi^*\bar{f}$ on $\Gamma_C$ extended arbitrarily to $\Gamma$ whenever
it is useful. $\Pi^*\bar{f}$ is gauge invariant (i.e., constant along each null direction),
and we call it a (weak) Dirac observable. 
In the relational observables literature, the Dirac observable corresponding to such a function $\bar{f}_\tau$ is denoted by the symbol
\begin{equation} F_{[f,T]}(\tau)\ :=\ \Pi^*\bar{f}_\tau. \label{OfTtau}\end{equation}
In this way, given  a function $f$ on $\Gamma$ and a system $T$ of clock functions $T^m$, one defines for every family $\tau$ of numbers $\tau^m$ the Dirac observable
(\ref{OfTtau}).

\subsubsection{Gauge transforming a relational Dirac observable}

Given a function $f:\Gamma_C\rightarrow \mathbb{R}$, the dependence on $\tau$ of the Dirac observable $F_{[f,T]}(\tau)$ can be interpreted as an evolution, 
\begin{equation}\label{evol1}  \tau\ \mapsto\ F_{[f,T]}(\tau), \end{equation}
or in terms of the physical phase space, as an evolution of the  physical observable,
\begin{equation}\label{evol2}  \tau\ \mapsto\ \bar{f}_\tau. \end{equation}
More generally, for every gauge transformation, that is, 
a map $\alpha:\Gamma_C\rightarrow \Gamma_C$ which preserves each null leaf, we can define a transformed gauge invariant function: 
\begin{equation} \label{gauge}\alpha\ \mapsto\ F_{[\alpha^*f,T]}(\tau). \end{equation}
In this framework \cite{Bianca1} one restricts to the gauge transformations preserving the foliation of $\Gamma_C$ defined by the clock system $T$, which is such that
\begin{equation}\label{tau'} \alpha^* T^m\ =\ T^m+{\tau'}^m,\ \ \ \ \ \ \  {\tau'}^m\in\mathbb{R},m \in {\cal M}. 
\end{equation}
Then, the gauge transformation amounts to
\begin{equation}\label{gauge'}  \tau' \ \mapsto\ F_{[f,T]}(\tau+\tau'). \end{equation}
Therefore, this definition does not really add to Eq. (\ref{evol1}) anything  more; however, we keep it for the consistency with \cite{Bianca1}.

\subsubsection{Dirac bracket}

One more element bridging  the kinematical phase space $\Gamma$ with the  physical phase space $\bar{\Gamma}$ with the help of the system $T$ of the clock functions is the Dirac bracket,
$$C(\Gamma)\ni f,g \ \mapsto \ \{f,g\}^*\in C(\Gamma),  $$ 
which may be defined using Eq. (\ref{defsigmatau}) by the following equality:
$$ \overline{\{f,g\}^*}_\tau\ =\ \{\bar{f}_\tau,\bar{g}_\tau\}_{\rm phys} $$
which is required to be true for every $\tau$. 
The relational construction of observables $F_{[\cdot,\cdot]}(\cdot)$, the Dirac bracket $\{\cdot,\cdot\}^*$, the Poisson bracket in $\Gamma$, and the ``gauge transformations'' (\ref{gauge'}) are consistent in the following way \cite{Bianca1}:
\begin{equation}\label{{F,F}}
\{F_{[f,T]}(\tau+\tau'),F_{[g,T]}(\tau+\tau')\}\ =\ F_{[\{f,g\}^*,T]}(\tau+\tau')
\end{equation}
[of course $\tau'$ could be erased without any loss of information, but
we keep it for the consistency with  Ref. \cite{Bianca1} and Eq. (8.20) therein].

\subsubsection{Incorrect statements\label{secincorrect}}

It is often  stated in the relational observables literature  about 
this action  of the gauge transformations (\ref{gauge}, \ref{gauge'})  and the evolution (\ref{evol1}, \ref{evol2}) that:
\begin{enumerate}
\item  They naturally pass to a map defined in the set ${\cal D}\subset C(\Gamma_C)$ of the gauge invariant functions on $\Gamma_C$,
\begin{equation}\label{wronga}          
\hat{\alpha}\ :\ {\cal D}\ni F_{[f,T]}(\tau)\ \mapsto\ F_{[\alpha^*f,T]}(\tau) \in {\cal D}
\end{equation}
by a fixed gauge transformation $\alpha$,  or, respectively, to a map 
\begin{equation}\label{wrongb}
\hat{\alpha}:{\cal D}\ni F_{[f,T]}(\tau)\ \mapsto\ F_{[f,T]}(\tau+\tau') \in {\cal D},
\end{equation}    
where $\tau'$ is defined by Eq. (\ref{tau'}) (see Eq. (8.15) in Ref. \cite{Bianca1}).
%
\item For every  $\alpha$ or $\tau'$, the corresponding maps (\ref{wronga}, \ref{wrongb}) preserve the Poisson bracket $\{\cdot,\cdot\}$ restricted to ${\cal D}$ (see the interpretation of Eq. (8.20) in Ref. \cite{Bianca1}).
\end{enumerate}
Let us explain why item 1 above is not true. Given a function $F_{[f,T]}(\tau)$ on ${\Gamma}_C$,  the function $f$  is neither known nor  unique. The would-be evolution  (\ref{wronga}) would be  well-defined if the right-hand side of Eq. (\ref{wronga}) were independent of that ambiguity. Unfortunately, the right-hand side does depend on that choice; given the function $F_{[f,T]}(\tau)$, we can choose functions $f_1$ and $f_2$ such that
$$F_{[f_1,T]}(\tau)\  =\ F_{[f_2,T]}(\tau)\ = F_{[f,T]}(\tau);$$
nonetheless,
$$ F_{[f_1,T]}(\tau + \tau') \neq\  F_{[f_2,T]}(\tau + \tau').$$ 
The maps (\ref{wronga}, \ref{wrongb}) are therefore ill-defined.

Since item 1 above is not true, item 2 is pointless. But it is even worse than that. Itself, the consistency  Eq. (\ref{{F,F}})  is true, and one could think
that it will eventually imply 2 as soon as the definition 1 is fixed.      
In this paper we correct the idea of item 1 of the action of the gauge  transformations in ${\cal D}$. In a quite clear and natural way, we fix item 1. We identify the additional structures behind the resulting gauge transformations induced in ${\cal D}$, and we define and characterize all the possibilities. Ironically however, the resulting gauge transformations in general do not satisfy item 2.  
We find the conditions upon which 2 is satisfied  and alternative conditions upon which 2 is not satisfied. Obviously Eq. (\ref{{F,F}}) continues to be true regardless of the case we consider. What is wrong is  \emph{inferring} 2 from Eq. (\ref{{F,F}}).

\section{Corrected approach}

Our aim is to use the idea of (\ref{evol1}, \ref{evol2}) to define an evolution in the physical phase space $\bar{\Gamma}$ correctly. We just need to modify the idea suitably, 
so that it leads to a consistent definition. Therefore, we use here the structures and notation introduced in the previous section.

\subsubsection{Reference functions}

The weakness of the attempt to apply Eq. (\ref{evol2}) was an overcompleteness of the set of all the functions $f$ on $\Gamma_C$ [more precisely, of their restrictions to a slice $\sigma_\tau(\bar{\Gamma})\subset\Gamma_C$]. To cure it, we choose a functionally complete, but \emph{not overcomplete}, set of functions on $\Gamma_C$.  That is, we fix a system $\theta$ of functions $\theta^I, I\in {\cal I}$, defined in $\Gamma_C$, such that their restrictions to the slices  $\sigma_\tau(\bar{\Gamma})$ form a coordinate system for each value of $\tau$. 
In other words, the assumption is that the pullbacks  $\sigma_\tau^*\theta^I$ onto the physical phase space $\bar{\Gamma}$ provide a coordinate system. 
We will call $\theta$ a system of reference functions.

\subsubsection{Map $\bar{f}\mapsto f$}

With this additional structure $\theta$, for every function $\bar{f}$ on $\bar{\Gamma}$, we can define a unique function $f$ on $\Gamma_C$, which will be used in Eq. (\ref{evol1}). To do this, we write $\bar{f}$ in terms of the coordinates,
\begin{equation} \label{tildef} \bar{f}(\bar{\gamma})\ =\ \tilde{f}((\sigma_0^*\theta^I)(\bar{\gamma})), \ \ \ \ \ \bar{\gamma}\in\bar{\Gamma}, \end{equation}
where $\tilde{f}$ is thus defined uniquely. This leads to a unique extension  $f$ of the function $\bar{f}$ to $\Gamma_C$,
\begin{equation} f(\gamma)\ =\ \tilde{f}(\theta^I(\gamma)), \ \ \ \gamma\in\Gamma_C. \end{equation}
This is an ``extension'' in the sense that  the function $f$ coincides with  $\Pi^*\bar{f}$ on the slice $\sigma_0(\bar{\Gamma})$,
\begin{equation} \left.f\right|_{\sigma_0(\bar{\Gamma})}\ =\ \left.\Pi^*\bar{f}\right|_{\sigma_0(\bar{\Gamma})}. \end{equation}

\subsubsection{Relational evolution defined in $C(\bar{\Gamma})$}

Finally, with 
\begin{align} C(\bar{\Gamma}) &\rightarrow C(\Gamma_C),\\
\bar{f}\ &\mapsto\ f, \end{align}
every $\tau$ defines a map 
\begin{align}C(\bar{\Gamma})\ &\rightarrow\ C(\bar{\Gamma})\\ 
\bar{f}\ &\mapsto\ \bar{f}_\tau := \sigma^*_\tau f.\label{evol4}\end{align}
This last map is an automorphism of the associative (not Poisson) algebra 
of functions as long as $\sigma_\tau^*\theta^I$ are coordinates on $\bar{\Gamma}$. 
In particular, for $\tau=\tau_0$ given by $\tau_0^m=0$, $m\in{\cal M}$, the map is the identity:
\begin{equation} \bar{f}\mapsto \bar{f}_{\tau_0}\ =\ \bar{f}. \end{equation}
In this way, we have completed our definition of the anticipated evolution (\ref{wronga})
\begin{equation}  (\bar f,\tau) \mapsto \bar{f}_\tau \label{evol5}\end{equation} 
in the Poisson algebra $C(\bar{\Gamma})$ of functions on the physical phase space $\bar{\Gamma}$.

\subsubsection{Dependence of the evolution on the choice of $\theta$}

The question we should ask is what the map (\ref{evol5}) depends on.
Our construction uses the system $T$ of clock functions $T^m$ \emph{and} the system $\theta$ of reference functions $\theta^I$ (coordinates on the slices defined via $T$ as $T^m = \tau^m$). Now, recall that on $\Gamma_C$, we have gauge transformations, the maps $\alpha: \Gamma_C \rightarrow \Gamma_C$ preserving the leaves generated by the null directions. Clearly every two pairs $(T,\theta)$ and $(T',\theta')$ which are gauge-equivalent (i.e., such that $T' = \alpha^*(T)$ and $\theta' = \alpha^*(\theta)$ for a gauge transformation $\alpha$) define the same evolution (\ref{evol5}).\footnote{
To convince the reader of this, consider $\bar{f}$ in $C(\bar{\Gamma})$ and two associated functions in $C(\Gamma_C)$, $f$ and $f'$, such that $f = \tilde{f} \circ \theta$ and $f = \tilde{f'} \circ \theta'$, where the tilded functions are defined by Eq. (\ref{tildef}) as $\bar{f} = \tilde{f} \circ \theta \circ \sigma_0$ and $\bar{f} = \tilde{f'} \circ \theta' \circ \sigma_0'$. Then, saying that the evolution is the same is equivalent to saying that $\sigma_\tau^* f = \sigma_\tau'^* f'$, or $f \circ \sigma_\tau = f' \circ \sigma'_\tau$. Now, recalling that $\sigma_\tau = \alpha \circ \sigma_\tau'$ and $\theta'^I = \theta^I \circ \alpha$, we can write the following sequence of identities: $f' \circ \sigma_\tau' = \tilde{f'} \circ \theta' \circ \alpha^{-1} \circ \sigma_\tau = \tilde{f'} \circ \theta \circ \sigma_\tau = \tilde{f} \circ \theta \circ \sigma_\tau = f \circ \sigma_\tau$. In the second-to-last step, we used the fact that $\tilde{f'} = \tilde{f}$ coming from $\tilde{f} \circ \theta \circ \sigma_0 = \bar{f} = \tilde{f'} \circ \theta' \circ \sigma'_0 = \tilde{f'} \circ \theta \circ \sigma_0$.}
Ignoring possible topological nontrivialities, every two systems $T$ and $T'$ are gauge-equivalent. Therefore, without loss of generality, we can fix a clock function system $T_0$, and the freedom of choosing the reference function system $\theta$ is enough to construct all maps (\ref{evol5}).

\subsubsection{Relation between $\theta$ and $\frac{\partial}{\partial T}$}

We can do even better. Indeed, not all the information involved in $\theta$ is relevant for the evolution (\ref{evol5}). Given $\theta$, consider $\theta'$ such that
\begin{equation} \theta'^{I'}\ =\ {\tilde{\theta}}'^{I'}(\theta^I).\label{thetatheta'} \end{equation}
The evolution (\ref{evol5}) defined by $\theta'$ is the same as that defined
by $\theta$. 
The relevant part of the choice of $\theta$ is encoded in the vector fields $\frac{\partial}{\partial T^m}$ corresponding to the coordinate system set by the functions
$\theta^I,T^m$ on the constraint surface $\Gamma_C$. Conversely,
for a given system $T$ of clock functions, let us consider a set of vector fields
$\partial_m$ on $\Gamma_C$ satisfying
\begin{equation}\partial_m (T^n)=\delta_{m}^n,\qquad\left[\partial_m,\partial_n\right]=0.\end{equation}
Due to the Frobenius theorem, locally, there is  a local solution $\theta^I$ to the equations
$$ \partial_m\theta^I=0, \ \ \ \ \ \ \ \ I\in {\cal I}.$$
That is enough if we keep ignoring the global nontrivialities. In this case,
however, owing to Eq. (\ref{thetatheta'}), our definition of the evolution (\ref{evol4}) extends consistently from one local chart $\bar{\theta}_\tau$ to another local chart
$\overline{\theta'}_\tau^{I'}$.

\section{Symplectomorphicity conditions}

\subsubsection{(Non)preserving of $\{\cdot,\cdot\}_{\rm phys}$}

As shown, item 1 of Sec. \ref{secincorrect}  has been cured by the introduction of a system $\theta$ of reference functions.
What about item 2? Does the map (\ref{evol4}) preserve the physical Poisson bracket (\ref{physPoiss}) defined on $C(\bar{\Gamma})$? 
For every fixed family of numbers $\tau$, we have defined coordinates on $\bar{\Gamma}$, 
\begin{equation} \bar{\theta}^I_\tau\ :=\ \sigma_\tau^*\theta^I. \end{equation}
In terms of them, the evolution (\ref{evol4}) reads 
\begin{equation} \bar{\theta}^I_0\ \mapsto\ \bar{\theta}^I_\tau. \end{equation}
The 2-form $\Omega_C$ used in the previous section to define the physical Poisson bracket can be written as 
\begin{equation} \Omega_C\ =\ \Omega_{IJ}(\theta^K,T^n)d\theta^I\wedge d\theta^J + dT^m\wedge  (\omega_{mI}d\theta^I+\omega_{mn}dT^n). \label{OmegaC} \end{equation}
On the physical phase space $\bar{\Gamma}$, the formula for the physical symplectic form $\bar{\Omega}$ reads 
\begin{equation} \bar{\Omega}\ =\ {\Omega}_{IJ}(\bar{\theta}^K_\tau,\tau^n)d\bar{\theta}^I_\tau\wedge d\bar{\theta}_\tau^J .\label{barOmega} \end{equation}
The physical Poisson bracket between two functions $\bar{\theta}^I_\tau$ and $\bar{\theta}^J_\tau$
reads 
\begin{equation}\label{physPoisstheta} \{\bar{\theta}_\tau^I,\bar{\theta}_\tau^J\}_{\rm phys}\ =\ ({\Omega}^{-1})^{IJ}(\bar{\theta}^K_\tau,\tau^n). \end{equation}
Therefore, the Poisson bracket is preserved if and only if the 2-form $\Omega_C$ decomposed according to Eq. (\ref{OmegaC}) satisfies
\begin{equation}  \frac{\partial}{\partial T^n}{\Omega}_{IJ}(\theta^K, T^m)\ =\ 0. \label{sympmor}\end{equation}

Generically,  this condition is not satisfied, and the map (\ref{evol4}) is not a symplectomorphism; hence, item 2 is not true. In fact, in example 2 we will see that  every arbitrarily chosen family of maps $C(\bar{\Gamma})\rightarrow   C(\bar{\Gamma})$ labelled by $\tau$, 
$\bar{f}\mapsto \bar{f}_\tau$  can be obtained as the evolution (\ref{evol5}).

\subsubsection{Identities}

The condition $d\Omega_C=0$ implies conditions on the terms of the decomposition (\ref{OmegaC}) $\Omega_C$. In particular we have
\begin{equation}  d^{(\theta)}\left(\Omega_{IJ}d\theta^I\wedge d\theta^J\right)\ =\ 0  \label{dOmegaIJ}\end{equation}
where by $d^{(\theta)}$ we denoted the part of the exterior derivative involving only
the derivatives $\frac{\partial}{\partial \theta^I}$. This condition is equivalent to
simply
\begin{equation} d\bar{\Omega}\ =\ 0. \end{equation}
Another condition is 
\begin{equation}  \left(\frac{\partial}{\partial T^m}\Omega_{IJ}\right)d\theta^I\wedge d\theta^J\ =\ d^{(\theta)}\left( \omega_{mI}d\theta^I\right) \label{ddtOmegaIJ}\end{equation}

\subsubsection{Symplectomorphicity and Hamiltonians}

Therefore, a reference system $\theta$ and a clock system $T$ do satisfy the symplectomorphism condition (\ref{sympmor}) if and only if 
\begin{equation} d^{(\theta)}\left(\omega_{mI}d\theta^I\right)\ =\ 0. \label{domegam}\end{equation}
In this case, there are (locally) defined Hamiltonians $H_m$,
\begin{equation} \omega_{mI}d\theta^I\ =\ d^{(\theta)}H_m(\theta,T),\label{omehamdHm}\end{equation}
which project to $\tau$-dependent Hamiltonians on $\bar{\Gamma}$:
\begin{equation}\label{barH}
\bar{H}_m(\tau)\ =\ \sigma_\tau^*H_m, \ \ \ \ \ m\in{\cal M}.
\end{equation}
We will go back to this and specifically to how the corresponding Hamiltonian vector fields fit into this approach after the following three examples.

\section{Examples}

\subsubsection{Example 1: Trivial evolution}

Let $\bar{\theta}_I$ be coordinates on
$\bar{\Gamma}$. Define the system  $\theta$ of reference functions to consist 
of the functions
\begin{equation} \theta^I \ =\ \Pi^*\bar{\theta}^I. \end{equation}
In terms of those functions
\begin{equation} \Omega_C\ =\ \Omega_{IJ}(\theta^K)d\theta^I\wedge d\theta^J. \end{equation}
The resulting evolution is the identity
\begin{equation} \bar{\theta}^I_\tau\ =\ \bar{\theta}^I . \end{equation}
In this case we even did not need to fix on $\Gamma_C$ any system $T$ of clock functions. The result is independent of that choice.
\bigskip

\subsubsection{Example 2: Arbitrary evolution}

Suppose there is  given an arbitrary 
family of maps $C(\bar{\Gamma})\rightarrow C(\bar{\Gamma})$,
\begin{equation} \bar{f} \mapsto \bar{f}_\tau \label{arbitrary}\end{equation}
labelled  by the family $\tau$ of numbers $\tau^m$ such that:
\begin{itemize}
\item
 the family $\tau$ is consistent with a system $T$ of clock functions $T^m$ on $\Gamma_C$;
 \item for $\tau=0$, the map (\ref{arbitrary}) is the identity;
\item $\bar{f}_\tau(\bar{\gamma})$ is differentiable in each $\tau^m$ for every $\bar{\gamma}$ and every differentiable function $\bar{f}$.   
\end{itemize}
 We construct now
a system $\theta$ of reference functions $\theta^I$ on $\Gamma_C$, such that the corresponding evolution (\ref{evol5}) will coincide with 
$$ (\bar{f},\tau)\ \mapsto\ \bar{f}_\tau $$
given by Eq. (\ref{arbitrary}).

Let $\bar{\theta}^I$ be coordinates in $\bar{\Gamma}$. The map (\ref{arbitrary}) maps
each of them appropriately:
\begin{equation} \bar{\theta}^I\mapsto \bar{\theta}^I_\tau. \end{equation}
The suitable reference functions are defined at each point $\gamma\in\Gamma_C$ as follows: 
\begin{equation} 
\theta^I(\gamma)\ :=\ \left.\bar{\theta}^I_\tau(\Pi(\gamma))\right|_{\tau=T(\gamma)}. 
\end{equation}    
\bigskip

\subsubsection{Example 3: Standard choice}

In practice, we have at our disposal  the  auxiliary kinematical 
phase space $\Gamma$ endowed with  coordinate system $(p_\chi,q^\chi)$, 
where $\chi$ runs over a labelling set ${\cal X}$, such that the symplectic
form takes the canonical form   
\begin{equation} \Omega\ =\ \sum_{\chi}dp_\chi\wedge dq^\chi. \end{equation}
Suppose that on the constraint surface $\Gamma_C$ the coordinates 
\begin{equation} p_m, \ \ \ \ \ m\in {\cal M}\subset{\cal X}
\end{equation}
are determined by the remaining coordinates. Split the set of coordinates accordingly: 
\begin{equation} p_\chi\ =\ p_i,\,p_m \ \ \ \ \ \ \ \ q^\chi\ =\ q^i,\,q^m. \end{equation}
Hence,
\begin{equation} \left.p_m\right|_{\Gamma_C} \ =\ H_m, \ \ \ \ \ m\in {\cal M} \end{equation}
where $H_m:\Gamma\rightarrow \mathbb{R}$ is for every $m\in{\cal M}$ a function such that
$$ \partial_{p_{n}}H_m\ =\ 0,\ \ \ \ n\in {\cal M}.$$ 
Suppose that the index $m$ ranging the subset ${\cal M}$  labels also the null directions tangent to  $\Gamma_C$. 
This is what happens when  $\Gamma_C$ is defined by first class constraints. 
Choose for a system of clock functions
\begin{equation} T^m\ =\ q^m. \end{equation}
Let us choose for reference functions 
\begin{equation} \theta^I\ =\ q^i, p_j. \end{equation}
The 2-form $\Omega_C$ expressed by those functions is  
\begin{equation} \Omega_C\ =\ dp_i\wedge dq^i + dT^m\wedge dH_m. \end{equation}
Remarkably, somewhat for free, it satisfies the condition (\ref{sympmor})
for the corresponding evolution  in $\bar{\Gamma}$ to be Hamiltonian.
The corresponding Hamiltonians are the functions $ p_m $ restricted to slices 
of $\Gamma_C$ such that
$$ T^{m'}\ =\ {\rm const},\ \ \ \ \ \ \ m'\in{\cal M}.$$     

\bigskip

This method is often used in practice (e.g., Refs. \cite{Bianca3,DLP}) due to its simplicity
and efficiency.

\subsubsection{Example 4: Nontrivial clock function}

The starting point is similar to the previous example, namely 
\begin{equation}\Omega \ =\ dp\wedge dq + dP\wedge dQ\end{equation}
and $\Gamma_C$ is defined by the equation
\begin{equation}P-h(q,p)\ =\ 0. 
\end{equation}
Let the clock function
$$T=\tilde{T}(q,p,Q,P), $$ 
be defined in the whole $\Gamma$,  and, conversely, 
 $$ Q\ =\ \tilde{Q}(q,p,T,P).$$
 Finally, let 
 $$\theta^I\ = q,p. $$
Then, in the new coordinates $q,p,T,P$, 
$$\Omega_C\ =\  (1+h_{,p}\tilde{Q}_{,q}\ -\ h_{,q}\tilde{Q}_{,p})dp\wedge dq\ +\ \tilde{Q}_{,T}dh\wedge dT.$$
The symplectomorphicity condition 
$$ \left(h_{,p}\tilde{Q}_{,q}\ -\ h_{,q}\tilde{Q}_{,p}\right)_{,T}\ =\ 0,$$
generically is not satisfied.

\subsubsection{Message}

We learn from the examples that our relational evolution can be trivial (example 1) or  arbitrary (example 2). If the gauge fixing functions are just some of the coordinates and for the physical degrees of freedom we choose another subset of the coordinate system in which the symplectic 2-form had the canonical form, then,  in a case of first class constraints, the corresponding evolution is symplectomorphic (example 3). Finally,   
if in example 3 we make a nontrivial choice of clock function but leave the coordinates parametrizing the physical degrees of freedom, then, generically, the symplectomorphicity condition will be violated (example 4).

\section{From $(T, \theta)$ to the generators of evolution}

\subsubsection{Generators}

Given a reference system $\theta$ and a clock system $T$ on the constraint surface $\Gamma_C$, the corresponding evolution defined by the relational framework is generated just by the vector fields $\frac{\partial}{\partial T^m}$, $m\in {\cal M}$. What is nontrivial about those vector fields are their ($\tau$-depending) projections onto the physical phase space $\bar{\Gamma}$. They can be calculated by using the decomposition of $\partial_{T^m}$ into the null part and the part
tangent to the slices 
\begin{equation} T^{m'}\ =\ \tau^{m'},\ \ \ \ \ \ \ \ \ \ \ m'\in{\cal M}, \end{equation}
that is, by
\begin{equation} \partial_{T^m}\ =\ \partial_{T^m}-X_m\ +\ X_m, \end{equation}
such that
\begin{equation}\left(\partial_{T^m}-X_m\right)\righthalfcup \Omega_C\ =\ 0,\ \ \ \ \ \ \ X_m\ =\ X_m^I\partial_{\theta^I}.\label{mullandtang} \end{equation}
The evolution defined by $(\theta,T)$ in $C(\bar{\Gamma})$ is generated by the $\tau$-dependent vector fields on $\bar{\Gamma}$ labelled by $m\in {\cal M}$,
\begin{equation} 
\bar{X}_m(\tau)\ =\ \left.\Pi_*\partial_{T^m}\right|_{T=\tau}\ =\ \left.\Pi_*X_m\right|_{T=\tau}\ =\ X_m^I(\tau)\partial_{\bar{\theta}^I_\tau}. 
\end{equation}

\subsubsection{Calculation of $\bar{X}_m$}

The vector field $X_m$, can be calculated from Eqs. (\ref{OmegaC}, \ref{mullandtang}) by inverting the equality 
\begin{equation} X_m^I\Omega_{IJ}\ =\ \omega_{mJ} \end{equation} 
(notice that $\Omega_{IJ}$ is necessarily invertible).

\subsubsection{Additional conditions}

There are additional consistency conditions on $\Omega_{IJ}, \omega_{mI}$, and $\omega_{mn}$, namely Eq.
(\ref{mullandtang}) implies 
\begin{equation}\omega_{nn}\ =\ 0, \ \ \ \ \ \omega_{mn}-\omega_{nm}\ =\ X_n\righthalfcup\omega_m \end{equation}
and $d\Omega_C=0$ in addition to Eqs. (\ref{dOmegaIJ}, \ref{ddtOmegaIJ}) imply
\begin{equation} \frac{\partial}{\partial T^m}\omega_{nI}- \frac{\partial}{\partial T^n}\omega_{mI}\ -\ 2\frac{\partial}{\partial \theta^I}\omega_{[mn]}\ =\ 0\ =\ \omega_{[mm',m'']}. 
\end{equation}

\subsubsection{Symplectomorphisms once again}

Now, in the case (\ref{omehamdHm}) considered before when the corresponding evolution $C(\bar{\Gamma})\rightarrow C(\bar{\Gamma})$ preserved the Poisson bracket,
we have
\begin{equation} \bar{X}_m(\tau)\righthalfcup \bar{\Omega}\ =\ \omega_{mI}(T=\tau)d\bar{\theta}_\tau^I\ =\ d\bar{H}_m(\tau), \end{equation}
where the functions $\bar{H}_m$ were defined on $\bar{\Gamma}$ in Eq. (\ref{barH}).   
Hence, indeed, in this case $\bar{X}_m(\tau)$ are Hamiltonian vector fields, and
the Hamiltonians are the functions we had already defined before.

\section{Summary}

We have fixed the inconsistent definition of a relational  
evolution of the Dirac observables of Hamiltonian-constrained systems [Ref. \cite{Bianca1}, Eq. (8.15)]. 
To this end, in addition to a system, say, $T$ of clock functions used in the relational observables framework, we introduced a system of reference functions, say, $\theta$. With this structure,  the gauge transformations preserving the constant-value surfaces of the clock functions do induce a family of movements of the physical phase space $\bar{\Gamma}$. 
In general and in fact even generically, the movements are not symplectomorphisms of $\bar{\Gamma}$ (unlike what could be concluded from Ref. \cite{Bianca1}, Eq. (8.20), assuming that Ref. \cite{Bianca1}, Eq. (8.15) is true). While the induced movements are defined by pairs $(T,\theta)$, all the clock function systems $T$ are gauge equivalent to each other. Therefore, one can fix any system of clock functions on $\Gamma_C$ and vary only the reference function systems. The reference function systems which do induce symplectomorphisms in the physical phase space $\bar{\Gamma}$ are characterized by some special form taken by the 2-form $\Omega_C$. One can read off the corresponding Hamiltonians in that case.   On the other hand, every family of maps $\bar{\Gamma}\rightarrow \bar{\Gamma}$ (not necessarily symplectomorphic) labelled by
the labelling set of the clock function system can be obtained as the relational 
evolution corresponding to a suitable reference function system. One of the examples we illustrate our construction with shows in what way a naive choice of the clock and, respectively, reference systems provides a symplectomorphic relational
evolution. Another example shows that a nontrivial choice of a clock function system induces in the physical phase space a nonsymplectic evolution.

Our work provides the correctly defined evolution of Hamiltonian-constrained
systems. The issue of the physical evolution of those systems 
is still an outstanding problem of general relativity. We hope that our
correction of the relational approach will help to solve this problem. 
 
    
\section{Acknowledgements}
We benefitted from discussions with Alex Stottmeister, Kristina Giesel, Hanno Sahlmann, and Thomas Thiemann. This work was partially supported by the grant of Polish Ministerstwo Nauki i Szkolnictwa Wy\.zszego nr N N202 104838 and by the grant of Polish Narodowe Centrum Nauki nr 2011/02/A/ST2/00300.

\end{document}